\documentclass{arxiv}
\usepackage{amssymb}
\usepackage{epsfig}
\usepackage{subfigure}
\usepackage{cite}
\usepackage{rotating}
\usepackage{caption}

\title{ Benford's law predicted digit distribution of aggregated income taxes: the surprising conformity of Italian cities and regions
}
\author{Tariq A. MIR$^ {1,a}$,  Marcel AUSLOOS$^ {2,3,b}$
and Roy CERQUETI$^{4,c}$
 }
\address{
$^1$ Nuclear Research Laboratory, Astrophysical Sciences Division, Bhabha Atomic Research Centre, Srinagar-190 006, Jammu and Kashmir, India. \\$^a$email:  taarik.mir@gmail.com\\
$^2$eHumanities group, Royal
Netherlands Academy of Arts and Sciences, \\Joan Muyskenweg 25, 1096
CJ Amsterdam, The Netherlands. \\ $^3$GRAPES, Universit\' e de
Liege, Sart Tilman, B-4000 Liege, Belgium. Rue de la Belle
Jardiniere 483, B-4031, Angleur, Belgium. \\ $^b$email: 
marcel.ausloos@ulg.ac.be\\
 $^4$ Department of Economics and Law, \\University of Macerata, \\Via Crescimbeni, 20\\I - 62100  Macerata, Italy.\\ $^c$email:  roy.cerqueti@unimc.it  }

\begin{document}

\catchline{}{}{}{}{}

\maketitle


\section{Abstract}
The yearly aggregated tax income data of all, more than 8000, Italian municipalities are analyzed for a period of five years, from
2007 to 2011, to search for conformity  or not with  Benford's law, a counter-intuitive phenomenon observed in large tabulated data
where the occurrence of numbers having smaller initial digits is more favored than those with larger digits. This is done in
anticipation that large deviations from  Benford's law will be found in view of tax evasion supposedly being widespread across Italy.
Contrary to expectations, we show that the overall tax  income data for all these years is in excellent agreement with Benford's law.
Furthermore, we also analyze the data of Calabria, Campania and Sicily, the three Italian regions known for strong presence of
mafia, to see if there are any marked deviations from   Benford's law. Again, we find that all yearly data sets for Calabria and Sicily
agree with Benford's law whereas  only the 2007 and 2008 yearly data show departures from the law for Campania. These results are again
surprising in view of underground and illegal nature of economic activities of mafia which significantly contribute to tax evasion.
Some hypothesis for the found conformity is presented.
\section{Keywords}

aggregated tax income, Italy, Benford's law


\section{Introduction}	
In the present information age the collection, processing and dissemination of data from different financial activities, of utmost
importance for human welfare, is quite easy and convenient. However, for policy makers, the extraction of meaningful information, critical for plausible strategic decisions, from the sea of available data is a formidable challenge. The far reaching, adverse, consequences of using flawed data have been exemplified by the 2007 financial crisis in the initiation of which data of questionable quality being used in corporates and governments played a significant role \cite{Francis}. 

One statistical tool which can serve as a first check on the quality of (large) numerical data and thereby, to a great extent, simplifies
the deciphering of anomalies present in (large) data sets is the so-called Benford's law  \cite{Newcomb,Benford}. The law describes the counter-intuitive uneven distribution of numbers in large data sets. Usually,  it appears that the occurrence of the first digits of numbers has nothing to do with their abundance within the data. In fact, in a large data set,  the appearance of each digit from 1 to 9 as first digit is equally likely with a proportion of about 11\%. However, according to Benford's law, the appearance of digits is such that the distribution of the first digits tends to be logarithmic with numbers having smaller first digits appearing more frequently than those having larger first digits. Thus, the distribution is heavily skewed in favor of smaller digits with digits 1, 2 and 3 taking about 60\% of the total occurrences as first digits and the remaining six digits i.e. 4 to 9 left with only 40\% of the occurrences.

Benford's law was first reported by   Newcomb  \cite{Newcomb} following his observation that the pages of logarithmic table books get progressively cleaner as one moves from initial to latter pages, the first page being the dirtiest. About four decades latter, Benford rediscovered the phenomenon through a similar observation and established it on a more solid footing by testing its accuracy on a large volume of data he collected from diverse fields, e.g. physical constants, atomic and molecular masses, street addresses, length of rivers etc. and concluded that the occurrence of first significant digits follow a logarithmic distribution \cite{Benford}
\begin{equation}
P(d)= log_{10}(1+\frac{1}{d}), d= 1, 2, 3...,9
\end{equation}

where P(d) is the probability of a number having the first non-zero digit d and $log_{10}$ is logarithmic to base 10.\\
The theoretical proportions for each of the digits from 1 to 9 to be first significant digit are as shown in Table 1.
\begin{table}[h]
\tbl{The distribution of first significant digits as predicted by Benford's law}
{\begin{tabular}{@{}ccccccccccc@{}} \toprule
Digit \hphantom{00} & 1 & 2 & 3 & 4 & 5 & 6 & 7 & 8 & 9 \\
Proportion \hphantom{00} & 0.301 & 0.176 & 0.125 & 0.097 & 0.079 & 0.067 & 0.058 & 0.051 & 0.046 \\ \botrule
\end{tabular} \label{ta1}}
\end{table} 

The development of research on Benford's law into a full fledged field is as fascinating as the discovery of the law itself. Firstly, though the mathematical form of the law is very simple a complete mathematical understanding is yet to be achieved \cite{Berger,Canessa,MSHMSS182gauvrit08,MSHMSS186gauvrit09}. Furthermore, the law represents the only distribution of leading digits invariant under scale and base transformations\cite{Hill, Hill1, Pinkham, Pietronero}. Secondly, numerous data sets from diverse fields conform to the law \cite{Mir, Mir1, Mir2, Pain, Shao, Sambridge, Judge, Mebane, Roukema,Cho, Clippe,denHeijerAiben,Sandron,Leontsinis,BenfordMACHBI}. For an exhaustive
list refer to \cite{Online}. Yet, there is {\it a priori} no set of criteria to predict what type of data should conform to the law. Necessary and sufficient conditions are much in debate. Nevertheless,  (i) the presence of sufficient number of entries in the data, for digits to manifest themselves, (ii) spanning of several orders of magnitude, and (iii) the absence of any human restrictions, i.e. there are no built in minima or maxima, on the appearance of numbers in data are some properties that data under investigation must posses for conformity to the law\cite{Durtschi}.
\section{Relevant literature}
After the seminal work of Benford, the first digit phenomenon again came to prominence following the efforts of Nigrini who reported its frequent emergence in financial data \cite{Nigrini1, Nigrini, Nigrini2}. Furthermore, Nigrini provided the first practical application of Benford's law in the detection of tax evasion by hypothesizing that to save on their tax liabilities individuals might understate income items and overstate deduction items in their tax return files leading to overall distortion of digital frequencies. A Benford's law test successfully captured the manipulation of digital frequencies in the submitted data \cite{Nigrini1}. Furthermore, falsification of financial documents, manipulated trade invoices and tax returns submitted by companies have been clearly  unraveled \cite{Nigrini}. Today the law is routinely used by forensic analysts to detect error, incompleteness and wilfull manipulation of the financial data. The basic premise of the test is that first digits in real data, in general, have a tendency to approach the Benford distribution whereas people intending to play with the numbers, when unaware of the law, try to place the digits uniformly. Thus any departure from the law raises suspicion about the quality of the data \cite{Hill2} and/or in the process involved in its generation \cite{Varian}.

Benford's law has been successfully utilized to expose the intention of cheating in both corporates and the governments. In order to attract investment, firms must posses a strong financial basis. Moreover, to project a healthy, though sometimes superficial, picture they report enhanced profits and reduced losses which are not always the actual values. Such a manipulation of financial statements, known as cosmetic earning management, is achieved through the rounding of numbers. This was detected first for firms in New Zealand where the frequency of zeros and nines as second digit of reported earnings was respectively more and less than could be expected from Benford's law \cite{Carslaw}. Subsequently this unethical phenomenon has been reported to be the practice of the day for the firms worldwide \cite{Kinnunen}. A recent example of corporate data manipulation, on a truly global scale, is the 2011 LIBOR scandal in which a cartel of banks distorted the interest pricing process of the inter-bank loans \cite{Metz}. Countries, like firms, also falsify economic data when it is strategically advantageous \cite{Michalski}. Thus, questions have been raised about the data submitted by Greece to the Eurostat to meet the strict deficit criteria set by the European Union (EU). Among all the member states of EU, the data submitted by Greece has been found to have the greatest deviation from Benford's law \cite{Rauch}. Similarly, the macroeconomic data of China is a subject of much debate, as it has been alleged to be overstating its GDP numbers to mislead investors \cite{Holz}. Furthermore, a Benford's law based assessment of the macroeconomic data submitted by countries to the World Bank hints at deliberate falsification of the data from the developing countries \cite{Nye}.

The manipulative behavior percolates down to the local governments as recent studies using Benford's law have uncovered deficiencies in the data of municipalities and states of several countries. The cases in point are the Valejo City, Orange County and Jefferson County in U.S., whose local authorities have filed for bankruptcy. The digit distributions of the financial statements of all the three municipalities have been shown to have significant departures from that expected on the basis of Benford's law, thereby raising questions about the credibility of the statements on revenues and spending \cite{Haynes}. Further indications of data tampering by local governments have been uncovered through a Benford analysis of
the official financial reports of the fifty states of U.S. \cite{Johnson}. Another study from Brazil analyzing the digit distribution of 134,281 contracts issued by 20 management units in two states found significant deviations from Benford's law and concluded that there is a tendency to avoid conducting the bidding process and the rounding in determining the value of contracts \cite{Costa}.

In Italy, the municipalities represent the lowest level of the government responsible for providing services to the local residents. Some of these activities are property issues, such as building permits, street lump, garbage collection, public transport, etc., and social activities, such as child and elderly care services, etc. \cite{Bartolini}. Tax collection is a fundamental source of revenues for local governments enabling the efficient delivery of services \cite{Padovani}. On the other hand, tax evasion is known to be widespread across Italy with studies estimating
between quarter and half of the country's GDP to be hidden from authorities in the form of underground economy \cite{Schneider}. The evasion of taxes is detrimental to the financial health of municipalities. To contain any financial distress, the municipalities resort to scaling down of expenditure like cutting on the number of employees, reducing the salaries of those that continue to be employed and even complete stopping of some of the services \cite{Jones}. Thus, any financial distress of municipalities has severe repercussions on the lives of the taxpayers and municipal employees. It is important to have better oversight of the quality of financial statements and accountability over the use of funds. The concerns on data quality and the poor auditing procedures being used have resurfaced more vigorously following the bankruptcy of number of local government bodies across several industrialized countries during recent financial crisis.

In the present study, we analyze the yearly aggregated tax data of all the Italian municipalities (cities) for a period of five years from 2007 to 2011 to see if there are any deviations from Benford's law. Beforehand, given the scale of tax evasion in Italy, one expects that the digit distribution of tax data would be in complete disagreement of the predictions of Benford's law. However, we find that the tax data of all the Italian cities shows an impressive compliance to the law.

Furthermore, we also analyze the yearly (city cumulated) data from three Italian regions of Calabria, Campania and Sicily. The municipalities of these regions have, from common knowledge or assumption, a low level of governance, poor delivery of services, and substantial presence of mafia and organized crime \cite{Brosio,Calderoni}. Thus given the poor tax administration in the municipalities of these regions, we again anticipated to find some hints of tax evasion through deviations from Benford's law. Surprisingly, the data from these three regions also satisfy the law, except the years 2007 and 2008 data for Campania which show large deviations from Benford's law.

\section{Data analysis and Results}
\subsection{Data }
The Italian state is organized in four levels of government (i) a central government (ii) 20 regional governments (iii) 110 provincial governments and (iv) more than 8000 municipalities, at the time of this writing. The municipalities represent the lowest level of the government in the administrative structure of Italy. Each municipality belongs to one and only one province, and each province is contained in one and only one region. The total number of municipalities has slightly varied over the years. This is due to occasional administrative reorganization, through the acts of the Italian Parliament, leading to the creation of new municipalities and sometimes also the merger of two or more municipalities into one
\cite{IT}. Thus we have a total of 8101, 8094, 8094, 8092 and 8092 municipalities respectively for year 2007, 2008, 2009, 2010 and 2011, respectively. During this time interval,  7 municipalities have changed from a province to another one, - in  so doing also changing from a region (Marche)  to another (Emilia-Romagna), in 2008.

We have analyzed the yearly aggregated tax income (ATI) data of all municipalities for the period of five years from 2007 to 2011. The data has been obtained by (and from) the Research Center of the Italian Ministry of Finance and Economy (MFE).  We have disaggregated contributions at the  municipal level to the Italian GDP.

The standard methodology of a Benford analysis is to first count the appearances of each digit from 1 to 9 as the first digit of numbers in the data. Then the corresponding theoretical frequency of each digit as first digit is determined from Benford's law. This is followed by estimating the goodness of fit of the theoretical and observed digit distributions both graphically and also using suitable fitness test. These steps are explained through the analysis of the yearly tax data for all the Italian cities in Table 2. The $N_{Obs}$, the number of times each digit from 1 to 9 appears as the first significant digit in the corresponding data are shown for each yearly data set in columns 2, 4, 5, 7, 8 of Table 2. Also
shown are $N_{Ben}$, the corresponding counts (in brackets) for each digit as predicted by Benford's law:
\begin{equation}
N_{Ben}= N log_{10}(1+\frac{1}{d})
\end{equation} 
where $N$ for each year is the total number of records i.e. the number of municipalities. For example, the total number of municipalities for year 2007 is $N=8101$, as shown in column 2 of Table 2. The root mean square error ($\Delta{N}$)  is calculated from the binomial distribution \cite{Shao}
\begin{equation}
\Delta{N}= \sqrt{NP(d)(1-P(d))}
\end{equation} 
The observed count for digit 1 as first significant digit is 2433 for 2007, whereas the expected count from Benford's law is 2438.64 with an error of about 41.29. The expected count from Benford's law and the corresponding error depends only on the number of records in the sample of data. For the yearly ATI data the number of records for years 2008 and 2009 is 8094 and  is 8092  for years 2010 and 2011. Thus, the expected count from Benford's law and the corresponding errors are shown only once in Column 4 and 6 respectively for such cases. From a visual inspection of Table 2, it is found that the observed and expected digit distributions are in reasonable agreement within the margins of the calculated error. This is further illustrated in Fig. 1, where the observed proportion of the first digits are compared with those expected from Benford's law. Contrary to the {\it a priori} expectations, for the ATI data of all these years, the agreement  between the observed and theoretically predicted distributions is quite remarkable.

In view of the statistical quantification of the closeness between the observed and predicted digits distributions, the Pearson's $\chi^{2}$, the most widely used test in Benford's law literature, 
\begin{equation}
  \chi^{2}(n-1) =\sum_{i=1}^n\dfrac{(N_{Obs}-N_{Ben})^{2}}{N_{Ben}}
\end{equation}
is thereafter used. Of course, zero is not the significant digit when it occupies the extreme left of a number. Thus, in case of the first digit analysis,  we have $n=9$ "data points", whence $n-1=8$ degrees of freedom. The critical value of $\chi^{2}$, under $95\%$ confidence level, for acceptance or rejection of null hypothesis - the observed and theoretically predicted digit distribution are same, is $\chi^{2}(8)$=15.507. If the value of the calculated $\chi^{2}$ is less than the critical value,  then we accept the null hypothesis and conclude that the data fits Benford's
law.
\subsection{Results}
For the 2007 ATI  data (column 2 of Table 2), the $\chi^{2}$ is 15.36 (the last row and column 2 of Table 2), a value smaller than $\chi^{2}(8)=15.507$; whence the null hypothesis is accepted indicating in turn that the tax data for year 2007 follows Benford's law. The calculated $\chi^{2}$ for the year 2008 is 27.52 and for 2009 is 18.96 which are far greater than the critical value of 15.507. Therefore,  the null hypothesis must be rejected. Furthermore, the $\chi^{2}$ for year 2010 is 12.29 and for 2011 is 13.72 and both values are less than the critical value, such that the null hypothesis must be accepted.

From a visual examination of Fig. 1, showing the comparison of the observed proportion of the first digits with those expected from Benford's law, the conclusion is of an excellent compliance for all the years. Thus for the 2008 and 2009 ATI data, conclusions on compliance to Benford's law drawn from  Fig.1 and the $\chi^{2}$ test appear to be contradictory. However, it is known that the results of the $\chi^{2}$ test are sensitive to the number of records in the data under analysis. The rejection of the null hypothesis becomes difficult for samples of small sizes, called type
II error, whereas for large samples the test suffers from "excessive power", wherein even a small deviation from Benford's law turns out to be significant, called type I error. This leads to a wrongful rejection of the null hypothesis. The large data sets require increasingly better fits to pass the $\chi^{2}$ threshold for conformity \cite{Nigrini2}, although by inspection they give better fits than small data sets, and often fail a $\chi^{2}$ test that the small dataset passes. Thus, larger values of $\chi^{2}$ for the tax data of years 2008 and 2009, in our study,  are likely due to this excessive power, despite the fact that they visually appear to show clear and extraordinary conformity to Benford's law. In fact,
it is due to this excessive power that the calculated $\chi^{2}$ for 2007, 2010 and 2011 yearly data sets are only marginally smaller than the critical value of 15.507, though the close visual conformity, evident from  Fig. 1, compels one to expect much smaller values. Thus, rather than being an indication of departure of the ATI data from Benford's law,  the high  $\chi^{2}$ values  are a manifestation of the limitations of the Pearson's $\chi^{2}$ test itself. 

A second analysis is in order. It is widely accepted that the so-called underground or black economy is larger in the Southern regions of Italy than elsewhere. Studies have shown that personal income tax and value added tax evasion is highest in Calabria, Sicily and Campania \cite{Brosio}. In anticipation to find some support for this conjecture through the deviations from Benford's law, we specifically analyzed the data from these three regions. 

Calabria consists of 409 municipalities grouped in 5 provinces. We show the analysis for the tax data of all years for Calabria in Table 3. Again since the number of records is same for all the yearly data sets the Benford expected frequency of each digit and the corresponding error are same and are only shown once in Column 6 of the Table 3. The Pearson's $\chi^{2}$ are all much smaller than the critical value of 15.507. Therefore, we conclude that the  ATI data follows Benford's law, a fact which is also clearly attested by Fig. 2, where the observed and expected digit
distributions are compared.

The analysis for Sicily region is shown in Table 4. Again both the calculated $\chi^{2}$ and the corresponding graphical representation of Fig. 3 show an excellent compliance to the law.

The results for the Campania region are shown in Table 5. Here, the respective $\chi^{2}$ for 2007 and 2008 are large than the critical value for acceptance of null hypothesis. Thus, for these two years the ATI data of Campania region clearly deviates from Benford's law. The $\chi^{2}$ for years 2009 to 2011 are less than the critical value of 15.507, though only marginally. It may be noted here that unlike the case of 2008 and 2009 yearly data for all the municipalities, for which  the results of the $\chi^{2}$ test are contradictory to the inferences, evident from the corresponding figure Fig. 1, for Campania, the calculated $\chi^{2}$ and the observations from the corresponding figure Fig. 4 are complimentary.
The departure of Campania data from Benford's law can be clearly seen from Fig. 4 where the frequency of digit 3 is much less and that of digit 6 is much greater than expected from Benford's law.

\newpage
\begin{table}
\tbl{The Observed and Benford expected first digit distributions of Aggregated Income Tax data of all Italian cities. In brackets, the
expected count from Benford's law and the corresponding errors for the data on the column on the left.}
{\begin{tabular}{@{}llllllll@{}} \toprule
First Digit &  2007 & 2008 & 2009 & 2010  & 2011 & \\
\\ \colrule	
1  &    2433 (2438.64 $\pm$41.29) & 2421 &  2433 (2436.54 $\pm$41.27) & 2454 & 2471 (2435.93 $\pm$41.26)\\

2  &	1436 (1426.51 $\pm$34.28) & 1466 & 1472 (1425.27 $\pm$34.27) & 1459  & 1454 (1424.92 $\pm$34.26)\\

3  &	951  (1012.14 $\pm$29.76) & 932  & 918 (1011.26 $\pm$29.75) & 931  & 931 (1011.01 $\pm$29.74)\\
 
4  &	755 (785.07 $\pm$26.63) &	 756 	& 758 (784.39 $\pm$26.62) & 751  & 752 (784.20 $\pm$26.61)\\

5  &	635 (641.44 $\pm$24.3) &	 620 	& 637 (640.88 $\pm$24.29) & 645  & 628 (640.72 $\pm$24.29)\\

6  &	575 (542.36 $\pm$22.5) &	 578 	& 557 (541.89 $\pm$22.49) & 559  & 572 (541.76 $\pm$22.48)\\

7  &	478 (469.78 $\pm$21.04) &	 501 	& 510 (469.37 $\pm$21.03) & 499  & 505 (469.26 $\pm$21.02)\\

8  &	412 (414.37 $\pm$19.83) &	 384 	& 402 (414.01 $\pm$19.82) & 405  & 404 (413.91 $\pm$19.82)\\

9  &	426 (370.7 $\pm$18.81) &	 436 	& 407 (370.38 $\pm$18.8) & 389 & 380 (370.29 $\pm$18.8)\\
\botrule
\# records  &	8101 &	8094  &  8094  &  8092 & 8092\\
\botrule
\bf$\chi^{2}$ \hphantom{00} & \bf15.36 & \bf27.52  & \bf18.96  & \bf12.29 & \bf13.72 \\ \botrule
\end{tabular} \label{ta1}}
\end{table}

\begin{table}
\tbl{The Observed and Benford expected first digit distributions of ATI data of 409 municipalities of Calabria. In brackets, the expected count from Benford's law and the corresponding errors.}
{\begin{tabular}{@{}llllllll@{}} \toprule
First Digit &  2007 & 2008 & 2009 & 2010  & 2011 & \\
\\ \colrule
1  &    122  &   123    &  122   & 126  & 123 (123.12 $\pm$9.28)\\

2  &	62  &    67     &  67   & 68    & 68  (72.02 $\pm$7.7)\\

3  &	54  &    52     &  49   & 47    & 48  (51.1 $\pm$6.69) \\
 
4  &	40  &	 38 	&  41   & 42    & 39  (39.64 $\pm$5.98)\\

5  &	30  &	 30 	&  26   & 25    & 30  (32.38 $\pm$5.46)\\

6  &	30  &	 28 	&  31   & 33    & 35  (27.38 $\pm$5.05)\\

7  &	28  &	 30	&  29   & 30    & 26  (23.72 $\pm$4.73)\\

8  &	19  &	 17	&  20   & 21    & 22  (20.92 $\pm$4.46)\\

9  &	24  &	 24	&  24   & 17    & 18  (18.72 $\pm$4.23)\\
\botrule
\bf$\chi^{2}$ \hphantom{00} & \bf4.44 & \bf4.51  & \bf4.94   & \bf5.42 & \bf3.02 \\ \botrule
\end{tabular} \label{ta1}}
\end{table}

\begin{table}
\tbl{The observed and Benford expected  first digit distributions of ATI data of 390 municipalities of Sicily. In brackets, the expected count from Benford's law and the corresponding errors.}
{\begin{tabular}{@{}llllllll@{}} \toprule
First Digit &  2007 & 2008 & 2009 & 2010  & 2011 & \\
\\ \colrule
1  &    112  & 106  &  109  & 110 & 113 (117.4 $\pm$9.06)\\

2  &	76  & 80 & 79  & 77  & 76 (68.68 $\pm$7.52)\\

3  &	48  & 48  & 48  & 51  & 49 (48.73 $\pm$6.53) \\
 
4  &	30  &	 30 	& 30  & 29  & 31 (37.79 $\pm$5.84)\\

5  &	33  &	 32 	& 31  & 33  & 30 (30.88 $\pm$5.33)\\

6  &	31  &	 27 	& 28  & 22  & 26 (26.11 $\pm$4.94)\\

7  &	20  &	 26	& 25  & 31  & 27 (22.62 $\pm$4.62)\\

8  &	19  &	 17	& 20  & 15  & 15 (19.95 $\pm$4.35)\\

9  &	21  &	 24	& 20  & 22 & 23 (17.85 $\pm$4.13)\\
\botrule
\bf$\chi^{2}$ \hphantom{00} & \bf4.61 & \bf7.73  & \bf4.42   & \bf9.72
 & \bf5.76 \\ \botrule
\end{tabular} \label{ta1}}
\end{table}

\begin{table}
\tbl{The observed and Benford expected first digit distributions of ATI data of 551 municipalities of Campania. In brackets, the expected count from Benford's law and the corresponding errors.}
{\begin{tabular}{@{}llllllll@{}} \toprule
First Digit &  2007 & 2008 & 2009 & 2010  & 2011 & \\
\\ \colrule
1  &    170  &   173   &  171   & 171   & 171 (165.87 $\pm$10.77)\\

2  &	88  &    88     &  97   & 95    & 95 (97.03 $\pm$8.94)\\

3  &	53  &    51     &  50   & 50    & 48 (68.84 $\pm$7.76) \\
 
4  &	52  &	 55 	&  50   & 48    & 46 (53.4 $\pm$6.94)\\

5  &	37  &	 36 	&  39   & 45    & 48 (43.63 $\pm$6.34)\\

6  &	56  &	 55 	&  51   & 48    & 43 (36.89 $\pm$5.87)\\

7  &	28  &	 24	&  28   & 27    & 33 (31.95 $\pm$5.49)\\

8  &	30  &	 34	&  33   & 36    & 32 (28.18 $\pm$5.17)\\

9  &	37  &	 35	&  32   & 31    & 35 (25.21 $\pm$4.91)\\
\botrule
\bf$\chi^{2}$ \hphantom{00} & \bf21.65 & \bf23.02  & \bf14.56   & \bf13.56 & \bf13.34 \\ \botrule
\end{tabular} \label{ta1}}
\end{table}

\begin{figure}
\subfigure{\label{}\includegraphics[width=0.7\linewidth, height=1\linewidth,  angle=270, clip=]{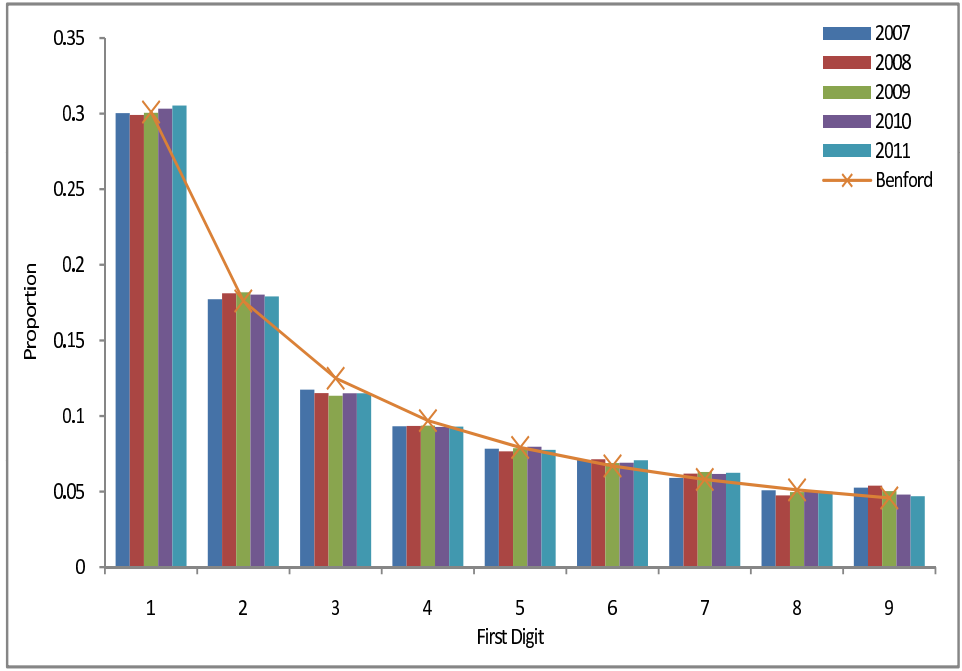}}
\vspace*{5pt}
\hspace*{0pt}
\caption{The observed and Benford expected first digit distributions of ATI data of all Italian cities.}
\vspace*{5pt}
\hspace*{20pt}
\end{figure}

\begin{figure}
\subfigure{\label{}\includegraphics[width=0.7\linewidth, height=1\linewidth,  angle=270, clip=]{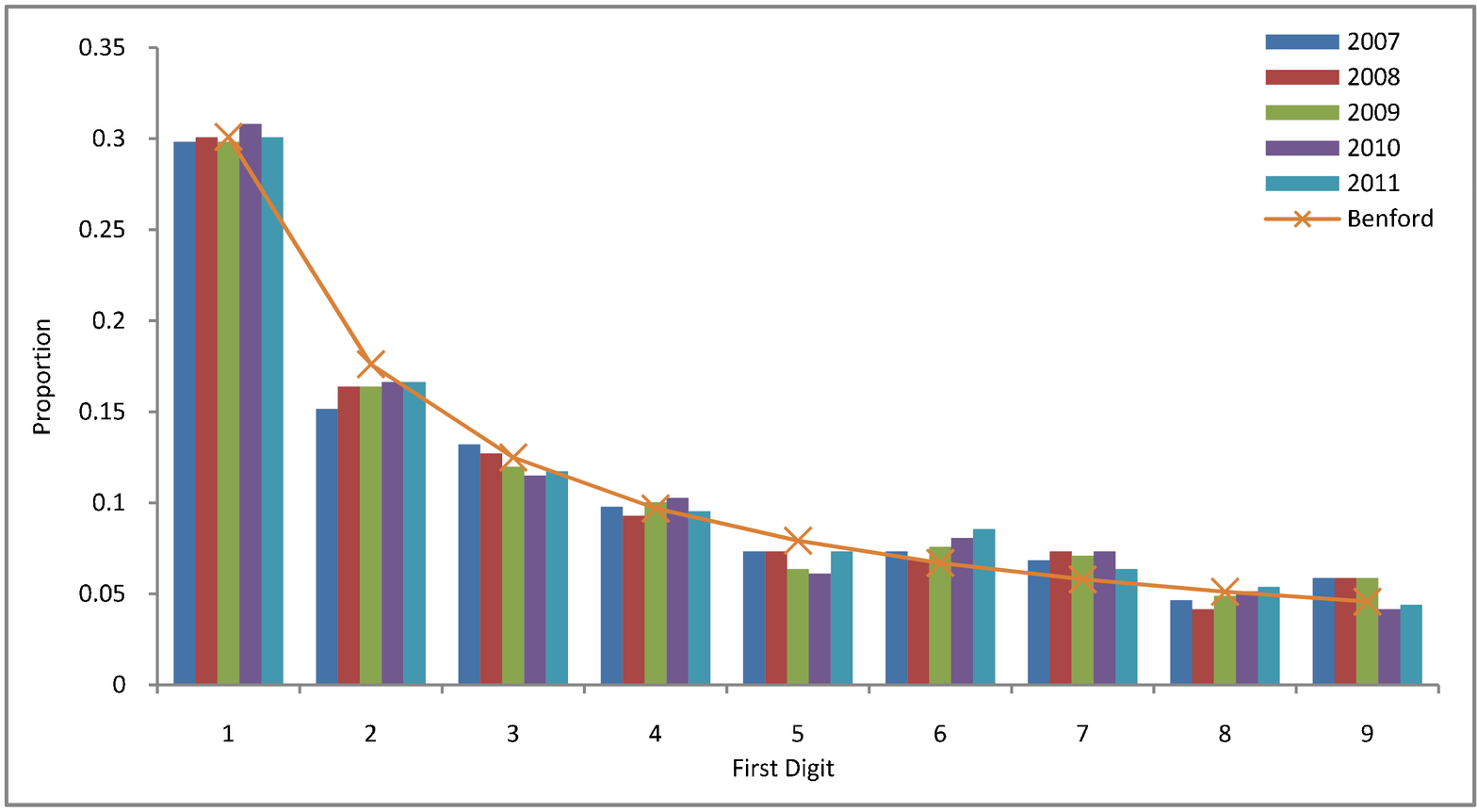}}
\vspace*{5pt}
\hspace*{20pt}
\caption{The observed and Benford expected  first digit distributions of ATI data of Calabria.}
\vspace*{5pt}
\hspace*{20pt}
\end{figure}

\begin{figure}
\subfigure{\label{}\includegraphics[width=0.7\linewidth, height=1\linewidth,  angle=270, clip=]{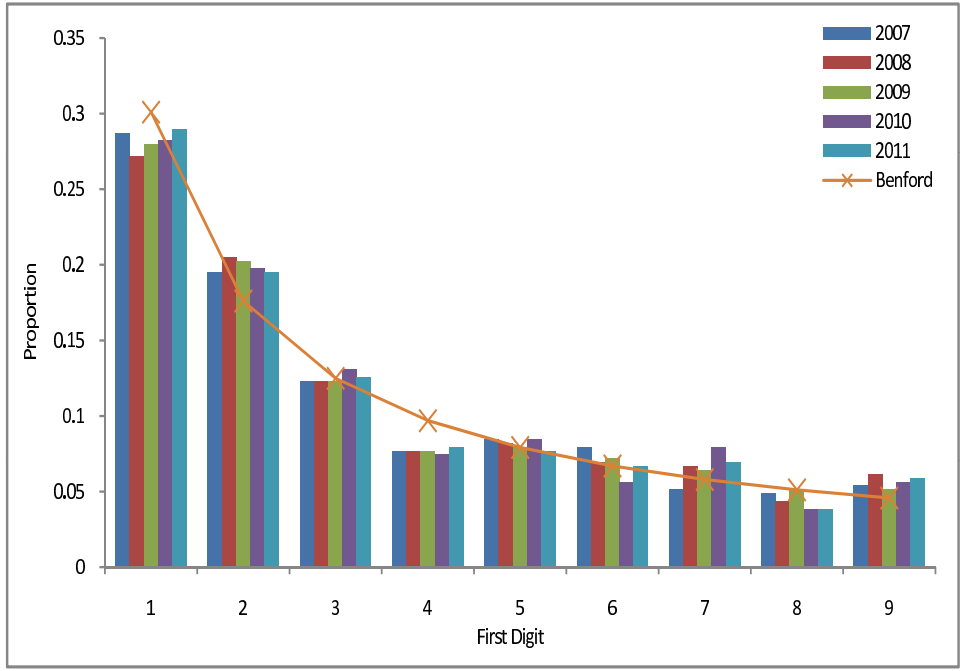}}
\vspace*{5pt}
\hspace*{20pt}
\caption{The observed and Benford expected  first digit distributions of ATI data of Sicily.}
\vspace*{5pt}
\hspace*{20pt}
\end{figure}

\begin{figure}
\subfigure{\label{}\includegraphics[width=0.7\linewidth, height=1\linewidth,  angle=270, clip=]{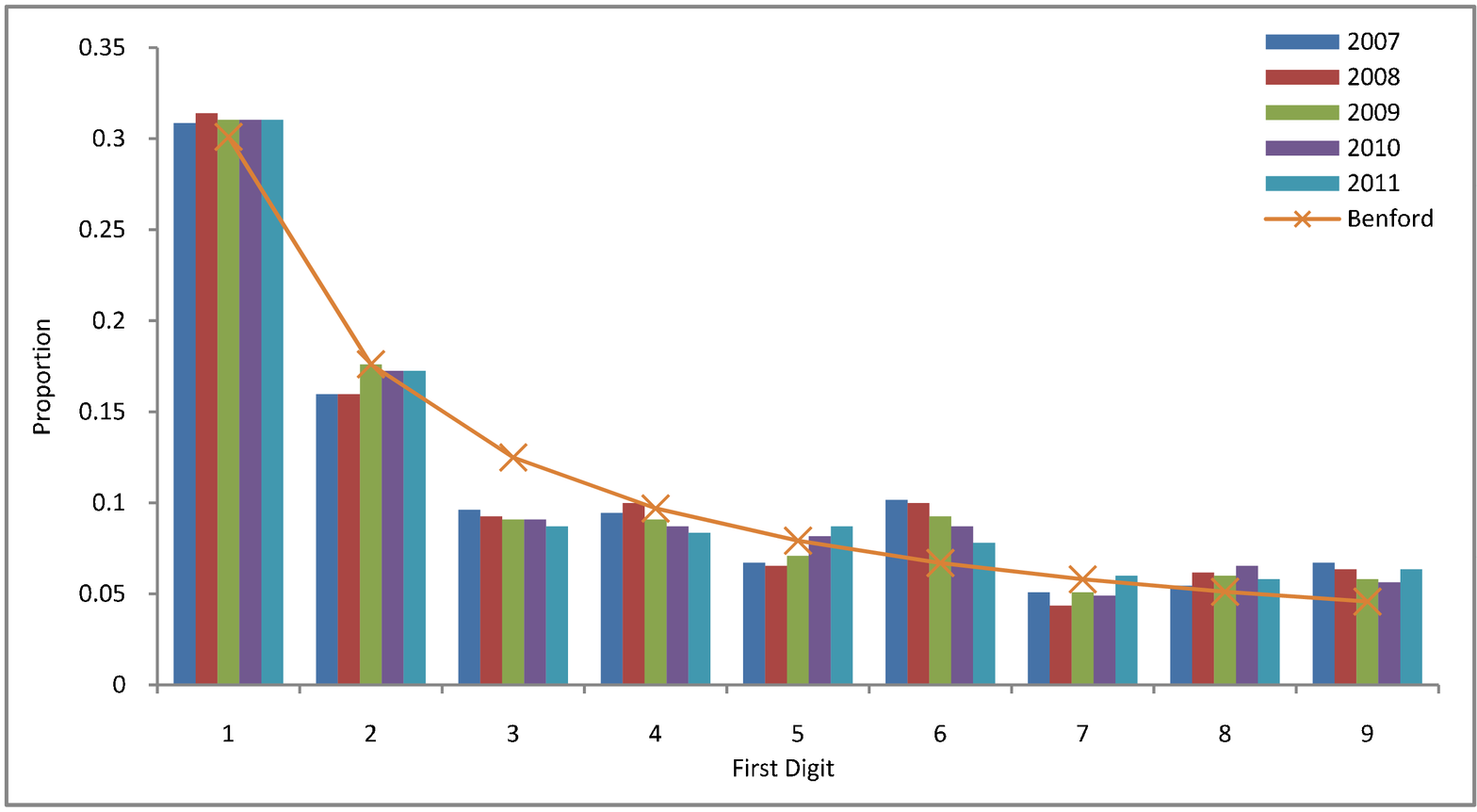}}
\vspace*{5pt}
\hspace*{20pt}
\caption{The observed and Benford expected  first digit distributions of ATI data of Campania.}
\vspace*{5pt}
\hspace*{20pt}
\end{figure}


\section{Discussion}
Though tax evasion in Italy is a phenomenon of gigantic proportions \cite{Schneider} its estimation is easier said than done \cite{Schneider1}. Nevertheless, the literature on estimation of tax evasion in Italy is plentiful. The levels of tax evasion have been shown to vary across different regions of Italy \cite{Brosio,Caro} and also  across different sections of the Italian society along its economic landscape \cite{Marino, Fiorio}. Furthermore, political and social factors responsible for  the thriving of tax evasion have been pointed out. For example, the lack of governance \cite{Brosio} and the levying of higher taxes are known to discourage taxpayers \cite{Bartolini}. The congestion of tax evaders is another factor encouraging evasion as tax authorities are overburdened thereby reducing the risk of detection \cite{Galbiati}.

The modeling of tax evasion is difficult as evasion in itself consists of partial or complete concealment of a significant proportion of economic  activities from the authorities. Thus researchers rely on sample surveys believing that people will disclose their true incomes when promised anonymity \cite{Fiorio}. The data from surveys are then compared with the official data maintained by Italian MFE. The latter is in fact the primary source of data invariably employed in modeling the different aspects of tax evasion. However, there is some general concern and much skepticism on the quality of tax data being maintained by MFE. The existence of several inconsistencies related to format, syntax and semantic have
been pointed out. More serious, from the point of view of the subject in this present study, are the missing, obsolete, or incorrect data values and undiscussed outliers \cite{Missier}. 

The present study assesses the quality of the MFE tax data relating to municipalities through the application of Benford's law in order to see  whether  there are in the ATI set, numerical  anomalies which might hint at deliberate attempts of manipulation. Any manipulation, if found, would in turn likely signal the intention of tax evasion. When people resort to under-reporting their taxable incomes, the occurrence pattern of digits in the tax return files will be altered in a major way. This has been shown to be true by Nigrini, in the first ever practical application of Benford's law, after analyzing tax return files of individuals on U.S. Internal Revenue Services. It was revealed that tax payers with low
incomes resorted to blatant manipulation of line items at filing time to a greater extent than high income tax payers \cite{Nigrini1}. Furthermore, prior studies suggest that Benford's law successfully captures the manipulation of tax data  to the point that the deviations from the law are acceptable as evidence in courts of several USA states. 

Our present study shows that overall tax data of Italian municipalities are in complete agreement with Benford's law, thereby negating the presence of evasion. At first, these results are somewhat surprising, since large scale presence of tax evasion in Italy is well documented.  However, on a more close scrutiny the reason for the compliance of data in our study becomes apparent. In our analysis, each municipality reports only one record of tax data for a particular year (in contrast to Nigrini analysis  of the data records of individual tax payers). Since municipalities are composed
of hundreds and thousands of inhabitants the tax value is an aggregate value which, of course, is a result of combinations of tax receipts of a large number of citizens. Thus, any individual deviation from Benford's law might be lost due to the multiple mathematical operations. In fact, the numbers obtained after multiplying a lot of numbers together have been found to agree with Benford's law. This has been conjectured to be one of the possible mathematical explanations for the validity of the law \cite{Pietronero,Boyle,Scott}.

It is relevant to mention here that another type of evasion comprises of complete concealment of incomes from the authorities. Such incomes usually arise from underground illegal activities like thefts, drugs, kickbacks, skimming and contract rigging. Since no record/trace of transactions exists at all the detection of tax evasion through such "underground" economic activities cannot be achieved with Benford's law.

Previous studies have shown that personal income tax and value added tax evasion is highest in Calabria, Sicily and Campania i.e. the Southern regions of Italy. The relative backwardness of these regions may be one reason for the high tax evasion since inefficiency of the municipalities in delivering services discourages the payment of taxes as tax payers do not see any proper return  for their paid taxes. The weak governments of poorer regions are generally less efficient in tax administration \cite{Brosio} further affecting the realization of tax revenues due to the strong
presence of black economy, arising out of  illegal and underground activities of mafia, (always) hidden from the authorities. Furthermore, extortion by mafia compels legal businesses to evade taxes. Fearful of their extensive reach and the deadly consequences for not obeying them businesses pay taxes to mafia rather than to government \cite{Alexeev}. Furthermore, the mafia infiltration of local governments of municipalities across these regions are pervasive and in order to restore the law enforcement against such infiltration the central government from time to time has resorted to the dissolution of the municipal administration. Over the years there have been a total 217 dissolution of local municipalities
across Italy. However, most of the dissolutions have been invoked in the municipalities of South-Italian regions with only 4 dissolutions being reported outside these regions \cite{Geys}. In the light of the above discussion, there are sufficient reasons to suspect that the tax data of these regions would show large departures from Benford's law. On the contrary, we found that income tax data from these regions shows strong submission to Benford's law except  for the 2007 and 2008 yearly data of  Campania. This is somewhat surprising since the Calabrian 'Ndrangheta' is the most powerful amongst all the Italian mafias \cite{Sergi}.
\section{Conclusion}
We have analyzed the yearly aggregated tax income data of all the Italian municipalities to search whether there are  anomalies which might hint at deliberate attempts of manipulation for tax evasion, a phenomenon widespread across the country, according to common knowledge. However, the overall data showed excellent compliance with Benford's law, thereby negating the presence of manipulations at this aggregate level. It might be that the aggregation process hides some individual breakdowns. However, the aggregation is not a multiplicative process.

We have also analyzed the municipality tax data of three regions (Calabria, Campania and Sicily) known for the strong presence of mafia. Again the data showed compliance to Benford's law except for the 2007 and 2008 data for Campania which showed significant departures from the law. Our findings suggests to reconsider the Campania data.

No need to say that the other (17) regions might be similarly investigated. Moreover, one possibility for further probing the data and maybe concluding on some reason for the breakdown of Benford's law, in a few cases, would be to investigate  the province level. The reader will easily understand that this demands 110 or so tests/year. This activity goes beyond the scope of the present report and is left open for further research.

Up to now, necessary and sufficient conditions for the application, and the more so, explanation of  Benford's law are not known, in spite of intense applications. The present report indicates that more theoretical and numerical investigations are still of interest.

\section*{Acknowledgments}

This paper is part of Ausloos's scientific activities in COST Action COST Action IS1104, "The EU in the new complex geography of economic systems: models, tools and policy evaluation" and in  COST Action TD1210 "Analyzing the dynamics of information and knowledge landscapes'"



\begin{thebibliography}{0}
\bibitem{Francis} L. Francis, V. R. Prevosto, Data and disaster: the role of data in the financial crisis, available at

$www.casact.org/pubs/forum/10spforum/Francis_{-}Prevosto.pdf$.
\bibitem{Newcomb} S. Newcomb, Note on the frequency of use of different digits in natural numbers, {\text Am. J. Math}. 4 (1881) 39-40.
\bibitem{Benford} F. Benford, The law of anomalous numbers, {\text Proc. Am. Phil. Soc.} 78 (1938) 551-572.
\bibitem{Berger} A. Berger, T. P. Hill, Benford's law strikes back: no simple explanation in sight for mathematical gem, {\text The Mathematical Intelligencer}, 33 (1) (2011) 85-91.
\bibitem{Canessa} E. Canessa, Theory of analogous force on number sets,
 Physica A  328 (1) (2003)   44--52.
\bibitem{MSHMSS182gauvrit08} N. Gauvrit, J-P. Delahaye, Pourquoi la loi de Benford n'est pas myst\' erieuse,  Math. \& Sci. Hum. / Math.  Soc. Sci.  182 (2008) 7-15.
\bibitem{MSHMSS186gauvrit09} N. Gauvrit, J-P. Delahaye, Loi de Benford g\' en\' erale, Math. \& Sci. Hum. / Math.  Soc. Sci.  186 (2009)   5-15.
\bibitem{Hill} T. P. Hill, Base-invariance implies Benford's law, {\text Proc. Am. Math. Soc.} 123 (3) (1995) 887-895.
\bibitem{Hill1} T. P. Hill, A statistical derivation of the significant-digit law, {\text Stat. Sci.} 10 (4)  (1995) 354-363.
\bibitem{Pinkham}  R. S. Pinkham, On the Distribution of First Significant Digits, {\text The Annals of Mathematical Statistics}, 32(4) (1961) 1223-1230.
\bibitem{Pietronero} L. Pietronero, E. Tosatti, V. Tosatti, A. Vespignani, Explaining the uneven distribution of numbers in nature: the laws of Benford and Zipf, {\text Physica A} 293 (2001) 297-304.
\bibitem{Mir} T. A. Mir, The law of the leading digits and the world religions, {\text Physica A} 391 (2012) 792-798.
\bibitem{Mir1} T. A. Mir, The Benford law behavior the religious activity data, {\text Physica A} 408 (2014) 1-9.
\bibitem{Mir2} T. A. Mir, The leading digit distribution of worldwide Illicit Financial Flows, $arXiv:1201.3432$
\bibitem{Pain} J. C. Pain, Benford's law and complex atomic spectra, {\text Phys. Rev. E.} 77 (2008) 012102.
\bibitem{Shao} L. Shao, B. Q. Ma, First digit distribution of hadron full width, {\text Mod. Phys. Lett. A} 24 (2009) 3275-3282.
\bibitem{Sambridge} M. Sambridge, H. Tkal\u{c}i\'{c}, A. Jackson, Benford's law in the natural sciences, {\text Geo. Phys. Res. Lett. A} 37 (2010) L22301.
\bibitem{Judge} G. Judge, L. Schechter, Detecting problems in survey data using Benford's law, {\text Journal of Human Resources.} 44 (2009) 1-24.
\bibitem{Mebane} W. R. Mebane, Jr., The wrong man is president! Overvotes in the 2000 presidential election in Florida, {\text Presp. on Polit.} 2 (2004) 525-535.
\bibitem{Roukema} B. F.   Roukema, Benford's Law anomalies in the 2009 Iranian presidential election, $arxiv:0906.2789v3$.
\bibitem{Cho} W. K. T. Cho, B. J. Gaines, Breaking the (Benford) law: statistical fraud detection in campaign finance, {\text The American Statistician} 61 (2007) 218-223.
\bibitem{Clippe} P. Clippe, M. Ausloos, Benford's law and Theil transform of financial data, {\text Physica A} 15 (2012) 6556-6567
\bibitem{denHeijerAiben} E. den Heijer and A. E. Eiben. Using aesthetic measures to evolve art. In: {\text IEEE Congress on Evolutionary Computation (CEC 2010)}, Barcelona, Spain, 18-23 July 2010. IEEE Press.
\bibitem{Sandron} F. Sandron, Do Populations Conform to the Law of Anomalous Numbers? Population-E 57 (2002)  755-761.
\bibitem{Leontsinis} T. Alexopoulos, S. Leontsinis, Benford's Law and the Universe, In press Journal of Astronomy and Astrophysics (2014).
\bibitem{BenfordMACHBI} M. Ausloos, C. Herteliu, B. Ileanu, Breakdown of Benford's Law for  Birth Data. Submitted.
\bibitem{Online} T. P. Hill, A. Berger, Benford Online Bibliography, 

 $http://www.benfordonline.net/$
\bibitem{Durtschi} C. Durtschi, W. Hillison, C. Pacini, The effective use of Benford's law to assist in detecting fraud in accounting data, {\text Journal of Forensic Accounting} 1 (2004) 17-34.
\bibitem{Nigrini1} M. J. Nigrini, Taxpayer compliance application of Benford's law, {\text The J. Am. Tax. Assoc.} 18 (1) (1996) 72-92.
\bibitem{Nigrini} M. J. Nigrini, L. J. Mittermaier, The use of Benford's law as an aid in analytical procedures, {\text Auditing: J. Pract. Theory} 16 (2) (1997) 52-67.
\bibitem{Nigrini2} M. J. Nigrini,  Benford's law: applications for forensic accounting, auditing and fraud detection, New Jersey, USA: Wiley Publications (2012).
\bibitem{Hill2} T. P. Hill, The difficulty of faking data, {\text Chance}  12 (3) (1999) 27-31.
\bibitem{Varian}  H. Varian,  Benford's law, {\text American Statistician} 23  65-66 (1972).
\bibitem{Carslaw} C. Carslaw, Anomalies in Income Numbers: Evidence of Goal Oriented Behavior, {\text Accounting Rev.} 63(2) (1988) 321-327.
\bibitem{Kinnunen} J. Kinnunen,  M. Koskela, Who is Miss World in cosmetic earnings management? An analysis of small upward rounding of net income numbers among 18 countries, J. Intern. Acc. Res. 2 (2003) 39-68.
\bibitem{Metz}  R.M. Abrantes-Metz, G. Judge, S. Villas-Boas, Tracking the Libor rate, {\text Applied Economics Letters}, 10(10) (2011) 893-899.
\bibitem{Michalski} T. Michalski, G. Stoltz, Do countries falsify economic data strategically? Some evidence that they might, {\text The Review of Economics and Statistics}, 95 (2013) 591-616.
\bibitem{Rauch} B. Rauch, M. G\"{o}ttsche, G. Br\"{a}hler, S. Engel, Fact and fiction in EU-Governmental economic data, {\text German Economic Review} 12 (2011) 243-255.
\bibitem{Holz} C. A. Holz, The Quality of China's GDP Statistics, Stanford University, SCID Working Paper 487, 2 December
2013.
\bibitem{Nye} J. Nye, C. Moul, The political economy of numbers: on the application of Benford's law to international macroeconomic statistics. {\text BE Journal of Macroeconomics}, 7(1),  article 17 (2007).
\bibitem{Haynes} A. H. Haynes, Detecting Fraud in Bankrupt Municipalities Using Benford's Law, {\text Scripps Senior Theses.}, Paper 42. (2012) available at 

$http://scholarship.claremont.edu/scripps_{-}theses/42$
\bibitem{Johnson}  G. G. Johnson, J. Weggenmann, Exploratory research applying Benford's law to selected balances in the financial statements of state govenments, {\text  Academy of Accounting and Financial Studies Journal}, 17(1) (2013) 31-44.
\bibitem{Costa} J. Costa, J. Santos,  S. Travassos, An Analysis of Federal Entities Compliance with Public Spending: Applying the Newcomb-Benford Law to the 1st and 2nd Digits of Spending in Two Brazilian States, R. Cont. Fin. - USP, Sao Paulo, 23(60), 187-198, set./out./nov./dez. 2012.
\bibitem{Bartolini} D. Bartolini, R. Santolini, Political yardstick competition among Italian municipalities on spending decisions, {\text The Annals of Regional Science}, 49(1) (2012) 213-235.
\bibitem{Padovani} E. Padovani, E. Scorsone, Measuring financial health of local governments a comparative framework, {\text Year Book of Swiss Administrative Sciences} (2011).
\bibitem{Schneider} F. Schneider, The increase of the size of the shadow economy of 18 OECD countries: some preliminary explanations, IFO Working Paper, 306 (2006).
\bibitem{Jones} G. Jones, Italy approves decree to stave off bankruptcy for Rome council, Feb. 28 (2014). available at

$http://www.reuters.com/assets/print?aid=USBREA1R1OD20140228$.
\bibitem{Brosio} G. Brosio, A. Cassone, R. Ricciuti, Tax evasion across Italy: rational noncompliance or inadequate civic concern, {\text Public Choice}, 112(3) (2002) 259-273.
\bibitem{Calderoni} F. Calderoni, Where is the mafia in Italy? Measuring the presence of the mafia across Italian provinces, {\text Global Crime}, 12(1) (2011) 41-69.
\bibitem{IT} $http://www.comuni-italiani.it/nomi/index.html$
\bibitem{Schneider1} F. Schneider, The value added of underground activities: size and measurement of the shadow economies of 110 countries all over the world, {\text  World Bank Working Paper} Washington, D.C. (2000).
\bibitem{Caro} P. Di Caro, G. Nicotra, Knowing the unknown across regions: spatial tax evasion across Italy,

$http://dx.doi.org/10.2139/ssrn.2446803$
\bibitem{Marino} M. R. Marino, R. Zizza, The personal income tax evasion in Italy: an estimate by taxpayer's type. In: Michael Pickhardt and Aloys Prinz (eds.), {\text Tax Evasion and the
Shadow Economy}, Cheltenham: Edward Elgar, 2011.
\bibitem{Fiorio} C. V. Fiorio, F. D'Amuri, Workers tax evasion in Italy, {\text Giornale degli Economisti e Annali di Economia}, 64 (2/3) (2005) 247-270.
\bibitem{Galbiati} R. Galbiati, Giulio Zanella, The tax evasion social multiplier: evidence from Italy, {\text Journal of Public Economics}, 96(5) (2012) 485-494.
\bibitem{Missier} P. Missier, G. Lalk, V. Verykios, F. Grillo, T. Lorusso, P. Angeletti, Improving data quality in practice: a case study in the Italian public administration, {\text Distributed and Parallel Databases}, 13(2) (2003) 135-160.
 \bibitem{Boyle} J. Boyle,
 An application of Fourier series to the most significant digit problem,  Am. Math. Month. 101 (1994) 879-886.
 \bibitem{Scott} P. D. Scott, M. Fasli, Benford's law: An empirical investigation and a novel explanation, {\text CSM Technical Report 349},  CSM technical report, Department of Computer Science,
University Essex. Available at $repository.essex.ac.uk$, 2001.
\bibitem{Alexeev} M. Alexeev, E. Janeba, S. Osborne, Taxation and evasion in the presence of extortion by organized Crime, {\text J. Comparative Economics}, 32 (2004) 375-387.
\bibitem{Geys} B. Geys, G. Daniele, Organized Crime, Institutions and political quality: empirical evidence from Italian municipalities. Workshop Paper, Dept. of Political Science, Stanford University, available at

$https://politicalscience.stanford.edu/sites/default/files/workshop-materials/Paper_{-}Ginmarco_{-}0.pdf$
\bibitem{Sergi} M. A. Sergi, Ndrangheta and gangster politics in Calabria. The local side of a global threat. Unpublished manuscript availabe at
$ecpr.eu$.

\end{thebibliography}
\end{document}